\documentstyle[prl,aps,epsf,multicol]{revtex}
\begin{document}
\draft
\title{Buckling instability in type-II superconductors with strong
pinning}
\author{R.G.~Mints}
\address{School of Physics and Astronomy,\\
Raymond and Beverly Sackler Faculty of Exact Sciences, Tel Aviv
University, Tel Aviv 69978, Israel}
\author{E.H.~Brandt}
\address{Max-Planck-Institut f\"{u}r Metallforschung, D-70569
Stuttgart, Germany}
\date{\today}
\maketitle
\begin{abstract}
We predict a novel buckling instability in the critical state of thin
type-II superconductors with strong pinning. This elastic instability
appears in high perpendicular magnetic fields and may cause an almost
periodic series of flux jumps visible in the magnetization curve. As an
illustration we apply the obtained criteria to a long rectangular
strip.
\end{abstract}
\begin{multicols}{2}
\narrowtext
In high magnetic fields a noticeable deformation of superconductors
occurs in the critical state because of the magnetic force density
${\bf f}={\bf j}\times{\bf B}$, where ${\bf j}$ is the current density
and ${\bf B}$ is the magnetic field. This results in an anomalous
irreversible magnetostriction (``suprastriction''\cite{kron}) and shape
distortion\cite{joh1,joh2} of type-II superconductors with strong
pinning. Similar as in magnetic fluid dynamics\cite{llcm} the stress
tensor of a superconductor in a magnetic field includes an additional
term, the Maxwell stress tensor of the magnetic field with components
of order $B^2/\mu_0$. Since this is quadratic in $B$, the Maxwell
stress tensor in the critical state can be important for elasticity in
strong magnetic field \cite{joh1,joh2}. However, even in a field of
$10\,{\rm T}$ the value of $B^2/\mu_0$ is small compared to the Young
modulus $E$ of the material. We estimate the ratio
$B^2/\mu_0E\approx 10^{-3}$ for $B\approx 10\,$T and $E\approx 100\,$GPa
which is a typical value for YBa$_2$Cu$_3$O$_{7-x}$ high-temperature
superconductors \cite{ting}.
\par
The effect of the magnetic field on the elastic behavior may be much
higher if one considers bending of thin samples since the effective
elastic modulus for bending $\tilde{E}$ is much less than the Young
modulus. In particular, for a long rectangular strip of extension
$l\times w\times d$ ($l\gg w\gg d$) one has
$\tilde{E}\approx E(d/l)^2\ll E$. If for instance $d/l\approx 10^{-2}$
and $E\approx 100\,$GPa, then $B^2/\mu_0$ is of the order of the
effective bending modulus $\tilde{E}$ at $B\approx 3.5\,{\rm T}$.
\par
\begin{figure} 
\epsfclipon
\vskip .75\baselineskip
\epsfxsize=0.9\hsize
\centerline{
\vbox{
\epsffile{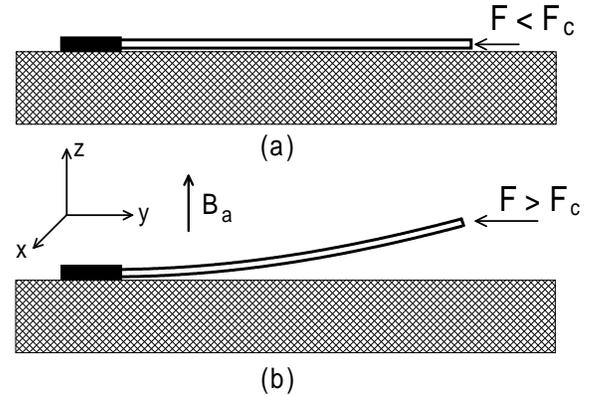}
}}
\vskip \baselineskip
\caption{Buckling of a thin superconductor strip with a clamped left
edge in a transverse magnetic field $B_a$.}
\label{fig1}
\end{figure}

An important consequence of a small value of the effective elastic
modulus for bending $\tilde{E}$ is the classical Euler buckling
instability \cite{llel,timo}. This elastic instability occurs for rods
and thin strips when the longitudinal compression force $F$ at the
edges of the sample exceeds a critical value $F_b\propto\tilde{E}$. In
particular, one has $F_b=\pi^2Ewd^3/48l^2$ for a long rectangular strip
with one edge clamped and the other edge free as shown in
Fig.~\ref{fig1} \cite{llel,timo}. The buckling instability manifests
itself at $F\ge F_b$ by a sudden bending with amplitude
$s\propto\sqrt{F-F_b}$.
\par
The magnetization of type-II superconductors with strong pinning and
the associated magnetic forces are successfully described by the Bean
critical state model\cite{bean} using a critical current density $j_c$
which decreases with increasing temperature and magnetic field. In the
transverse geometry of a thin strip in a perpendicular field one has
$j=j_c$ in the region where the magnetic flux has penetrated and
screening sheet currents $J$ with  $0 <J <j_c d$ in the flux-free
region \cite{ehb1,schu,ehb2}. This nonuniform flux distribution is not
in equilibrium and under certain conditions a thermomagnetic flux-jump
instability may occur producing a sudden intensive heat release. This
heat pulse decreases the critical current density and drives the system
towards the equilibrium state with a uniform flux. A sudden buckling of
a superconductor in the critical state may also lead to a heat pulse
and thus to a sudden flux penetration into the sample, which shows as a
flux jump instability in the magnetization curve.
\par
\par
In this letter we predict a novel Euler buckling instability caused by
the longitudinal magnetic compression force acting in the critical
state of a thin superconducting strip in a strong transverse magnetic
field. We discuss several scenarios how the buckling instability
develops, including the cases when a sudden buckling shows as a flux
jump instability in the magnetization curve. A series of buckling
induced flux jumps almost periodic in an increasing applied magnetic
field is predicted.
\par
\vskip -.5\baselineskip
\begin{figure} 
\epsfclipon
\narrowtext
\epsfxsize=0.9\hsize
\vskip .5\baselineskip
\centerline{
\vbox{
\epsffile{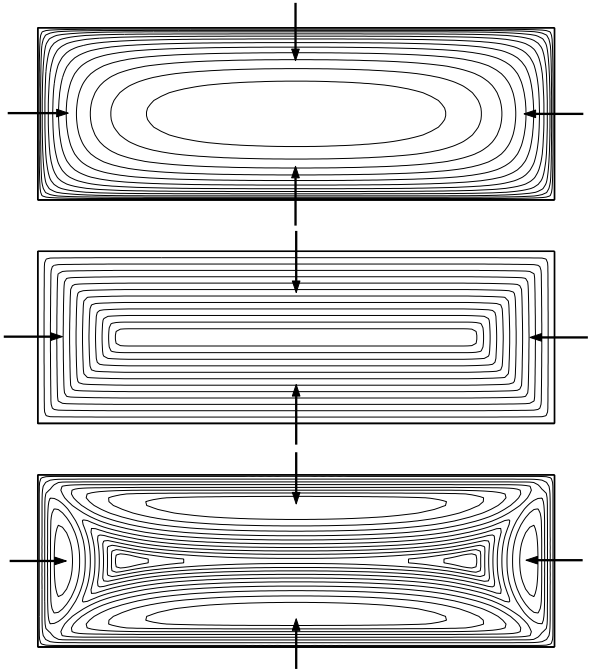}
}}
\vskip 1.5\baselineskip
\caption{The current streamlines in the critical state of a type-II
superconductor thin strip in a transverse magnetic field computed by
the method [9]. The arrows indicate the magnetic forces acting on the
strip. Top: Meissner state, $B_a\ll B_c$. Middle: Fully penetrated
critical state, $B_a\gg B_c$. Bottom: Applied field decreasing from
$B_0\gg B_c$ to $B_0-2.4 B_c$, which yields penetrating fronts with
inverse flux at $|x|=a/\cosh 2.4\approx 0.56a$ and a negative force
$F$, Eq.~(\ref{eq4}).}
\label{fig2}
\end{figure}
\par
We consider first the elastic stability of a long rectangular strip
$l\times w\times d$ ($l\gg w\gg d$) in an increasing transverse
magnetic field ${\bf B}_a \| \hat{\bf z}$, assuming that the strip is
glued to the substrate at the left edge ($y=0$) as shown in
Fig.~\ref{fig1}. A longitudinal compression force $F$ acts near the
right edge of the strip ($y=l$) in the area where the electromagnetic
force density ${\bf f}={\bf j}\times{\bf B}$ has a $y$-component due to
the U-turning current, thus
\begin{equation}   
F=-F_y=B_ad\!\int\!\!\!\int\! j_x\,dxdy,
\label{eq1}
\end{equation}
where the integral is over the U-turn area. As shown in
Fig.~\ref{fig2}, in the fully penetrated critical state this area is a
triangle where $j_y=j_c$, but in general the integral is over the right
half of the strip ($y>l/2$). If $w\ll l$ the deformation of the strip
can be obtained assuming that $F$ is applied to the very end of the
strip at $y=l$. For such narrow strips one can show that exactly
$F=B_aM$, where $M$ is the total magnetic moment of the strip divided
by its length $l$.
\par
Depending on the magnetic prehistory of the sample the dependence of
the longitudinal compression force $F$ on $B_a$ is described by the
following three formulas\cite{ehb2,ehb3}.
\par
(a) For a zero-field cooled straight strip (Fig.~1a) with $B_a$
increasing from zero one has
\begin{equation}  
F=j_cB_ada^2\tanh {B_a\over B_c} \,,
\label{eq2}
\end{equation}
where we introduce $a=w/2$ and $B_c=\mu_0j_cd/\pi$. The longitudinal
force $F(B_a)$, Eq.~(\ref{eq2}), has the limits
$F\approx \pi B_a^2a^2/\mu_0$ ($B_a\ll B_c$, Meissner state) and
$F\approx j_cB_ada^2$ ($B_a\gg B_c$, fully penetrated critical state).
\par
(b) For $B_a$ {\it increasing} from a field-cooled value $B_0$, one has
the force
\begin{equation}  
F=j_cB_ada^2\,\tanh{B_a-B_0\over B_c}\,.
\label{eq3}
\end{equation}
\par
(c) For $B_a$ {\it decreasing} from a fully penetrated critical state
with $B_a= B_0$, the force $F=B_a M$ decreases as
\begin{equation}  
F=j_cB_ada^2 \Big[\,1 - 2\tanh{B_0-B_a\over 2 B_c}\,\Big]\,,
\label{eq4}
\end{equation}
going through $F=0$ at $B_a\approx B_0 -1.1 B_c$.
For a narrow strip the field of full penetration is \cite{ehb3}
\begin{equation}  
B_p=B_c\,\Big(1+\ln\!{w\over d}\,\Big).
\label{eq5}
\end{equation}
\par
In the case of a curved strip (Fig.~1b) in the formulae (2-4) for $F$
the factor $B_a = F/M$ means the $z$ component of $B_a$, while in the
argument of $\tanh (\dots)$ the $B_a$ should be replaced by the
component $B_\perp$ of $B_a$ perpendicular to the strip near its right
end (where the U-turning currents flow). In general, the magnetic
moment $M$ and the force $F=B_a M$ depend on the prehistory of
$B_\perp(t)$ and may relax with time $t$.
\par
If the buckling instability for a zero-field cooled strip occurs at a
field $B_b > B_p$, the force is
\begin{equation}  
F=j_cda^2B_b.
\label{eq6}
\end{equation}
The critical force $F_b$ for the buckling instability of a strip with
one edge clamped and the other edge free is \cite{llel,timo}
\begin{equation}  
F_b={\pi^2ad^3E\over 24l^2}.
\label{eq7}
\end{equation}
Equating the forces $F$ and $F_b$ we find that the magnetic field $B_b$
at which the first buckling instability occurs is
\begin{equation}   
B_b ={F_b \over j_c d a^2}
    ={\pi^2 \over 24}{E\over j_c a}\Big({d\over l}\Big)^2.
\label{eq8}
\end{equation}
\par
We estimate the fields $B_c\approx 0.04\,$T, $B_p\approx 0.15\,$T, and
$B_b\approx 4\,$T using the data for YBa$_2$Cu$_3$O$_{7-x}$
superconductors $E\approx 10^2\,$GPa \cite{ting}, and assuming that
$j_c\approx 10^9\,$A/m\,$^2$, $w\approx 10^{-3}\,$m,
$d\approx 10^{-4}\,$m, and $d/l\approx 10^{-2}$. This estimate verifies
our initial suggestion that $B_p\ll B_b$.
\par
The height $s$ of the right end of the buckled strip (see
Fig.~\ref{fig1}b) can be found analytically if the maximum angle
$\theta_{m}$ between the tangent to the buckled strip and the substrate
is small \cite{llel,timo}. Assuming that the force $F$ slightly exceeds
the critical value $F_b$ we obtain a sinusoidal bending with the
amplitude
\begin{equation}   
{s\over l}\approx {4\sqrt{2}\over\pi}\,\sqrt{{F\over F_b}-1}
\approx 1.8\,\sqrt{{F\over F_b}-1}
\label{eq9}
\end{equation}
and
\begin{equation}   
\theta_m\approx 2\sqrt{2}\sqrt{{F\over F_b}-1}=
{\pi\over 2}\,{s\over l}.
\label{eq10}
\end{equation}
\par
Now assume that the external magnetic field is increased with constant
ramp rate $\dot B_a$ and the threshold of the buckling instability is
reached when $F=F_b$. One can consider several scenarios how the
buckling evolves, depending on the value of $\dot B_a$ and on the ratio
of the time constants for bending of the strip, $\tau_b$, for magnetic
flux diffusion, $\tau_m$, and for heat diffusion, $\tau_h$, see
Ref.~\cite{rgm1} for details.
\par
The first scenario applies to a very low ramp rate $\dot{B}_a\ll
B_a/\tau_m$, where the current and magnetic field distributions inside
the strip, and thus the magnetic forces, follow the increasing field
$B_a$ without delay. In this case the strip starts to bend as soon as
the magnetic compression force $F$ reaches the critical value $F_b$.
The force $F=B_a M$ via the magnetic moment $M$ depends on the
perpendicular field component near the tilted tip of the long strip,
$B_\perp =B_a \cos\theta_m$. This means that in $\theta_m(F)$,
Eq.~(\ref{eq10}), $F$ depends on $\theta_m$ and one has to find the
value of $\theta_m$ self-consistently. To do this we need the
appropriate dependence $M(B_\perp)$. We shall see that the resulting
$B_\perp$ {\it decreases} with increasing $B_a$ (or time); thus we have
to use Eq.~(\ref{eq4}) with $B_0=B_b$ (the field where buckling starts)
and with $B_a$ replaced by $B_\perp$ in $\tanh(\dots)$. Expanding the
hyperbolic tangent we thus find for $B_b -B_a \ll B_c$,
\begin{equation}  
F\approx F_b\,{B_a\over B_b}\,\Big( 1-{B_b-B_\perp\over B_c}\Big).
\label{eq11}
\end{equation}
Inserting this force into Eq.~(\ref{eq10}), $\theta_m^2 =8(F/F_b -1)$,
and solving for $\theta_m$ using $B_\perp\approx B_a(1-\theta_m^2/2)$
and $B_a \gg B_c$ we obtain
\begin{equation}  
\theta_m\approx \sqrt{2}\,\sqrt{{B_a\over B_b}-1}.
\label{eq12}
\end{equation}
\par
This self-consistent tilt angle $\theta_m$ is two times less than the
tilt angle  Eq.~(\ref{eq10}) for constant compression force $F$. The
physical origin of this negative feedback is the reduction of the total
U-turning current and thus of the force $F$, caused by the decrease of
$B_\perp$ when the end of the strip tilts, compare the current
distributions in Fig.~\ref{fig2}.
\par
A different scenario appears when the buckling occurs with a delay at a
force $F_d$ slightly above $F_b$ (``overheating"). Several reasons for
such a delay are conceivable, e.g., sticking of the strip to the
substrate by adhesion, or a misalignment of the perpendicular applied
magnetic field ${\bf B}_a$ such that the force ${\bf F}$ in
Fig.~\ref{fig1} points slightly downward to the substrate. A small
misalignment is probably inevitable for a typical experiment.
\par
When after zero-field cooling $F=F_d = j_c B_d da^2$ is reached at
$B_a=B_d$, the buckling amplitude jumps almost instantly to a finite
value $s\sim\sqrt{F_d/F_b -1}$. To obtain this amplitude
self-consistently one may combine Eq.~(\ref{eq10}) for $\theta_m(F)$
with Eq.~(\ref{eq4}) for $F(\theta_m)$, like in the first scenario,
noting that $M$ and thus the force $F = B_a M$ depend on
$B_\perp = B_a\cos\theta_m$. The sudden jump of $\theta_m$ at $B_a=B_d$
means that $B_\perp$ is reduced from $B_d$ to $B_d (1 -\theta_m^2 /2)$
(if $\theta_m^2 \ll 1$) and thus Eq.~(\ref{eq4}) is required yielding
\begin{equation}   
F = F_d \, \Big[\, 1 -2\tanh{\theta_m^2 B_d \over 4 B_c}\,\Big]
\label{eq13}
\end{equation}
with $F_d = j_c B_d da^2$. Inserting this into Eq.~(\ref{eq10})
and solving for $\theta_m^2 \ll 4B_c /B_d$ one obtains for
$B_d \gg B_c$:
\begin{equation}   
\theta_m^2={2B_c \over B_d}\Big({F_d \over F_b} -1 \Big)\,.
\label{eq14} 
\end{equation}
\par
Equation~(\ref{eq14}) differs from Eq.~(\ref{eq12}) because of the
different history of the perpendicular field $B_\perp(t)$ and thus of
the magnetic moment: In the first scenario $B_\perp$ started to
decrease from the lower threshold $B_b$ and the decrease occurs since
the rising $B_a$ is overcompensated by the growing $\theta_m$. In the
present scenario, $B_\perp$ has reached the higher threshold $B_d >B_b$
before it drops down, and this drop is solely due to the growing tilt
angle $\theta_m$ while $B_a =B_d$ is constant in this approximation. As
a consequence, the self-consistent tilt angle $\theta_m$ is reduced
much more in this case, by a factor $\sqrt{B_c/4B_d} \ll 1$.
\par
This strong feedback mechanism requires that the change of the current
density occurs {\it instantaneously}, much faster than the mechanical
buckling, $ \tau_m \ll \tau_b $. In reality the redistribution of the
currents will lag behind the buckling. In the extreme limit
$\tau_m \gg\tau_b$, the tilt angle would first jump to its original
large value $\theta_d^2= 8(F_d/F_b -1)$, Eq.~(\ref{eq10}), and then
relax to the small value of Eq.~(\ref{eq14}), or to zero, or to some
other value. The theoretical problem is intricate since a quantitative
treatment requires the self-consistent time dependent solution of the
equations for $B_\perp(t)$ with a relaxing, history dependent magnetic
moment $M\{B_\perp(t)\}$, using $B_\perp = B_a(t)(1-\theta_m^2/2)$ and
$\theta_m^2 = 8 (F/F_b -1)$ with $F=B_a M$. This yields the implicit
equation for $B_\perp(t)$,
\begin{equation}   
B_\perp(t)=B_a(t)[\,5 - 4B_a(t)M\{B_\perp(t)\}/F_b\,]\,,
\label{eq15} 
\end{equation}
from which the tilt angle $\theta_m^2(t)=2(B_\perp/B_a -1)$ is obtained.
To solve this one requires a realistic model for the relaxing history
dependent magnetization.
\par
From our numerical work we expect the magnetic relaxation to be very
fast and non-exponential when $\partial B_\perp /\partial t$ changes
sign \cite{ehb3}, as it is the case during buckling. During very fast
switching of $B_\perp(t)$, the electric field is so large that,
irrespective of pinning, the vortices exhibit usual flux-flow behavior,
with flux-flow resistivity $\rho_{\!f}\approx (B/B_{c2})\rho_n$, where
$B_{c2}$ is the upper
critical field and $\rho_n$ is the resistivity in the normal state. In
this case the magnetic relaxation time of an Ohmic strip applies,
$\tau_m \approx\tau_0 =0.249 ad\mu_0/\rho_{\!f}$ \cite{ehb4}. This time
has to be compared with the buckling time $\tau_b$, which we estimate
from the lowest resonance frequency $\omega_1$ of the strip (a
cantilevered reed \cite{ehb5}),
$\tau_b \approx \omega_1^{-1}$, $\omega_1^2\approx 1.03\,Ed^2/(\rho l^4)$
where $\rho$ is the specific weight. Inserting here numbers for
YBa$_2$Cu$_3$O$_{7-x}$ at $B_a = 4$ T, we estimate $\tau_m \ll \tau_b$,
i.e., the magnetic relaxation initially is instantaneous. With
proceeding relaxation, the electric field and the effective resistivity
decrease, and thus the magnetic relaxation time increases. We thus
expect that the real behavior of the strip is somewhere between the two
considered limits $\tau_m \ll \tau_b$ and $\tau_m \gg \tau_b$.
\par
Therefore, if buckling starts delayed at a force $F_d$ and disappears
at a smaller force $F_b < F_d$, the tilt angle at the tip of the strip
may oscillate between a maximum value  $\theta_{\rm max} \le \theta_d$
and zero. Such oscillation may occur since at $\theta_d$ the reduction
of $B_\perp$ is so large that the currents tend to change sign and thus
the force $F$ rapidly decreases. The tilt angle then may drop to zero,
undershooting the small equilibrium value, Eq.~(\ref{eq14}). With
continuously increasing applied field $B_a(t)$, the tilt angle
$\theta_m$ thus makes a sudden jump from zero to $\theta_{\rm max}$,
then drops rapidly back to zero, where it remains until the next
excursion occurs when $F$ again reaches $F_d$. These buckling
instabilities should occur at nearly equidistant field values with
period of the order of the penetration field $B_p$, Eq.~(\ref{eq5}),
and they will show up in the magnetization curve as a periodic set of
flux jumps.
\par
So far we assumed that the temperature $T$ of the strip stays constant,
$T=T_0$. However, buckling of a strip in the critical state causes some
heat release which increases the temperature and decreases the force
$F(T) \propto j_c(T)$. A complete solution of the buckling instability
in type-II superconductors with high critical current density should
therefore include a self-consistent treatment of the magnetic field and
temperature variations.
\par
For a rough estimate of the decrease of the force $F(T)$ we assume here
that $j_c(T)\propto (T_c-T)$, the critical temperature is $T_c\gg T_0$,
and the heating of the strip is adiabatic. In this case we find that a
sudden tilt to an angle $\theta_m$ leads to
\begin{equation}   
{F(T)-F(T_0)\over F(T_0)}\approx -{j_cwB_d\over
C(T_0)T_c}\,{\theta_m^2\over 2}.
\label{eq16}
\end{equation}
Combining Eqs.~(\ref{eq10}), (\ref{eq13}), and (\ref{eq16}) we find
that  self-heating affects the buckling instability threshold if
$C(T_0)T_c\ll j_cwB_c$, which results in $T_0 \lesssim 3\,$K for a
heat capacity $C(T)\approx 7\times T^3$J/Km$^3$ \cite{jano}.
\par
The temperature dependence of $F(T)$ may cause oscillations of the
strip. Indeed, a sudden buckling leads to a heat pulse increasing the
temperature $T$ and decreasing the force $F(T)$. If because of the
temperature increase the force $F(T)$ falls below the buckling
threshold $F_b$ then the strip straightens and the next instability
occurs after the strip has cooled down.
\par
In summary, we have shown that a strong magnetic field applied
perpendicular to a cantilevered superconductor strip will lead to Euler
buckling of this strip. We give the threshold field at which this
elastic instability occurs. During buckling, the effective applied
field at the tip of the strip decreases due to tilting. As a
consequence, the buckling force is reduced. This feedback mechanism may
lead to mechanical oscillations of the strip and its magnetization,
which depend on the magnetic and thermal relaxation times of the
specific experiment. At sufficiently low temperatures this sudden
buckling may trigger a periodic series of flux-jump instabilities which
should show in the magnetization curve.
\par
R.G.M. acknowledges numerous stimulating discussions with Dr.~A.~Gerber
and support from the Max-Planck-Institut f\"{u}r Metallforschung.
\par
\end{multicols}
\end{document}